\documentclass[final,1p,times]{elsarticle}
\usepackage{amssymb}
\usepackage{amsfonts}
\usepackage{amsmath}
\usepackage{geometry}
\usepackage{bm}
\numberwithin{equation}{section}
\newtheorem{thm}{Theorem}

\newdefinition{rmk}{Remark}
\newproof{pf}{Proof}
\newproof{pot}{Proof of Theorem \ref{thm2}}
\begin{document}

\begin{frontmatter}
\title{Generalization of the three-term recurrence formula and its applications}

\author{Yoon Seok Choun\corref{cor1}}
\ead{Yoon.Choun@baruch.cuny.edu; ychoun@gc.cuny.edu; ychoun@gmail.com}
\cortext[cor1]{Correspondence to: Baruch College, The City University of New York, Natural Science Department, A506, 17 Lexington Avenue, New York, NY 10010} 
\address{Baruch College, The City University of New York, Natural Science Department, A506, 17 Lexington Avenue, New York, NY 10010}
\begin{abstract}
The history of linear differential equations is over 350 years. By using Frobenius method and putting the power series expansion into linear differential equations, the recursive relation of coefficients starts to appear. There can be between two and infinity number of coefficients in the recurrence relation in the power series expansion. During this period mathematicians developed analytic solutions of only two term recursion relation in closed forms. Currently the analytic solution of three term recurrence relation is unknown.  In this paper I will generalize the three term recurrence relation in the linear differential equation. This paper is 2nd out of 10 in series ``Special functions and three term recurrence formula (3TRF).'' The next paper in series deals with the power series expansion in closed forms of Heun function \cite{Chou2012c}. The rest of the papers in the series show how to solve mathematical equations having three term recursion relations and go on producing the exact solutions of some of the well known special functions including: Mathieu, Heun, Biconfluent Heun and Lame equations. See section 7 for all the papers and short descriptions in the series. 
\end{abstract}

                                      
\end{frontmatter}

\section{\label{sec:level1}Introduction}
Mathieu functions\cite{Guti2003}, is an example of three term recurrence relation appears in physical problems involving elliptical shapes\cite{Troe1973} or periodic potentials, were introduced by Mathieu (1868)\cite{Math1868} when he investigated the vibrating elliptical drumhead. 

Mathieu function has been described using numerical approximations (Whittaker 1914\cite{Whit1914}, Frenkel and Portugal 2001\cite{Fren2001}).  Whittaker tried to obtain the analytic solution of Mathieu equation using Floquet's theorem and he reached the conclusion using three term recursive relation. He did not represent the solution of Mathieu equation in closed forms because of its three term recursive coefficients. He argued that ``While the general character of the solution from the function theory point of view is thus known, its [Mathieu] actual analytical determination presents great difficulties. The chief impediment is that the constant $\mu $ (the Mathieu exponent) cannot readily be found in terms of $a$ and $q$.''\cite{Whit1914}: $a$ and $q$ (see (1) in Ref.\cite{Whit1914}) are corresponding to $\lambda $ and $-\frac{1}{8} q$ (see (1) in Ref.\cite{Chou2012e}). 
In my opinion, Mathieu functions are difficult to represent in analytic closed forms and in its integral forms because of the three recursive coefficients.

Heun function, is an example of three term recurrence relation, generalizes all well-known special functions such as: Spheroidal Wave, Lame, Mathieu, and hypergeometric $_2F_1$, $_1F_1$ and $_0F_1$ functions. The Heun equation has four kinds of confluent forms such as Confluent Heun, Doubly-Confluent Heun, Biconfluent Heun and Triconfluent Heun equations. Mathieu equation is the special case of Confluent Heun equation.

 According to Hortacsu, `Heun differential equation\cite{Birk2007,Hort2007} starts to appear in astronomy or in general relativity\cite{Aliev1999} inevitably.  Heun equation is a general equation of all well-known special functions; Mathieu, Lame and Coulomb spheroidal equations. The power series expansion on Heun function can not be described as two term recursion relation any more. The coefficients in a power series expansions of Heun equation have a recursive relation between a 3-term.'\cite{Hortacsu:2011rr}

For the past 350 years, we only have constructed the power series expansion in closed forms using the two term recurrence relation in linear differential equation. However, ``since 1930, we do not have simple problems to solve in theoretical particle physics and scientists and mathematicians doing research on this field have to tackle more difficult problems, either with more difficult metrics or in higher dimensions. Most of the difficult problems must include three term or more'' (Hortacsu 2011 \cite{Hortacsu:2011rr}). 

In this paper I will construct three term recurrence relation in the form of power series expansion for the cases of polynomial and infinite series. In the following papers (see section 7) using the method described in this paper I will show how to obtain exact solution (1) in the power series expansion, (2) in the integral formalism and (3) in the generating function of any linear ordinary differential equations having three term recurrence relation. Using the method described in this paper one might be able to obtain exact solutions of higher term recurrence formula; four, five, $\cdots$, $m^{th}$, $\infty $ term recurrence formulas. If this is possible, then one would be able to generalize all homogeneous and inhomogeneous linear differential equations. 

Most problems in nature turns out to be nonlinear. For simplification purposes we usually try to linearize these nonlinear system. The systems are linearized using certain methods of simplification resulting in better approximation of physical systems. All physics theories (E \& M, Newtonian mechanics, quantum mechanic, QCD, supersymmetric field theories, string theories, general relativity, etc), generally involve solutions of linear differential equations. Unfortunately in some instances there are no analytic solutions to these physical systems. Hopefully the methods and procedures described in these papers (see section 7) will help obtain better analytic solutions to linearized nonlinear systems.
\section{\label{sec:level2}Lame equation, Frobenius method and three term recurrence relation}
There are no exact solutions in closed forms for second order (or higher order) ordinary  linear differential equations consisting of three term recurrence relation: some of the examples are Lame function, the generalized Lame function, Heun's equation, GCH function\cite{Ch2012}, Mathieu function, etc. Two term recurrence relation are solvable. Some examples are: Legendre function, hypergeometric function, Kummer function, Bessel function, etc. 
Let's think about an example which has no analytic solution in closed forms in detail.
\begin{equation}
\frac{d^2{y}}{d{t}^2} + \frac{1}{2}\left( \frac{1}{t-a} + \frac{1}{t-b} + \frac{1}{t-c} \right) \frac{d{y}}{d{t}} - \frac{\alpha (\alpha +1)t+\beta }{t(t-a)(t-b)(t-c)} y = 0
\label{eq:1}
\end{equation}
(\ref{eq:1})  is Lame differential equation. If $\alpha $ is not positive integer, the solution of it is called as the generalized Lame function. Replace $t$ by $x+a$ in (\ref{eq:1}). By using the function $y(x)$ as Frobenious series in it,
\begin{equation}
y(x) = \sum_{n=0}^{\infty } c_n (t-a)^{n+\lambda } = \sum_{n=0}^{\infty } c_n x^{n+\lambda }
\label{eq:2}
\end{equation}
$\lambda $ is an indicial root. Plug (\ref{eq:2}) in (\ref{eq:1}). And its recurrence relation is
\begin{equation}
c_{n+1}=A_n \;c_n +B_n \;c_{n-1} \hspace{1cm};n\geq 1
\label{eq:3}
\end{equation}
where,
\begin{subequations}
\begin{equation}
K_{n}=A_n  + \frac{B_n}{K_{n-1}} \hspace{1cm};n\geq 1
\label{eq:1.1}
\end{equation}
\begin{equation}
K_n =\frac{c_{n+1}}{c_n} \hspace{1cm} K_{n-1} =\frac{c_{n}}{c_{n-1}}
\label{eq:1.2}
\end{equation}
\begin{equation}
A_n = \frac{\{\alpha (\alpha +1)a+\beta  \}-2^2(2a-b-c)(n+\lambda )^2}{2^2(a-b)(a-c)(n+\lambda +1)(n+\lambda +\frac{1}{2})}
\label{eq:4.a}
\end{equation}
\begin{equation}
B_n = \frac{\{\alpha +2(n+\lambda )-1\} \{\alpha -2(n+\lambda )+2\}}{2^2(a-b)(a-c)(n+\lambda +1)(n+\lambda +\frac{1}{2})}
\label{eq:4.b}
\end{equation}
\end{subequations}
All other differential equations having no analytic solution in closed forms can be described as in (\ref{eq:3}). If I get a formula of (\ref{eq:3}) type, I can apply it to all other functions having no analytic solution in closed forms to diverse areas such as Lame function, generalized Lame function, Mathieu function, Heun function, GCH function\cite{Ch2012}, etc. 

\section{\label{sec:level3}Infinite series}
Assume that
\begin{equation}
c_1= A_0 \;c_0
\label{eq:5}
\end{equation}
(\ref{eq:5}) is a necessary boundary condition. The three term recurrence relation in all differential equations having no analytic solution in closed forms follow (\ref{eq:5}).
\begin{equation}
\prod _{n=a_i}^{a_i-1} B_n =1 \hspace{1cm} \mathrm{where}\;\mathrm{a_i}\;\mathrm{is}\;\mathrm{positive}\;\mathrm{integer}\;\mathrm{including}\;0
\label{eq:6}
\end{equation}
(\ref{eq:6}) is also a necessary condition. Every differential equations having no analytic solution in closed forms also take satisfied with (\ref{eq:6}). 

My definition of $B_{i,j,k,l}$ refer to $B_{i}B_{j}B_{k}B_{l}$. Also, $A_{i,j,k,l}$ refer to $A_{i}A_{j}A_{k}A_{l}$. For $n=0,1,2,3,\cdots $, (\ref{eq:3}) gives
\begin{equation}
\begin{tabular}{  l  }
  \vspace{2 mm}
  $c_0$ \\
  \vspace{2 mm}
  $c_1 = A_0 c_0 $ \\
  \vspace{2 mm}
  $c_2 = (B_1+ A_{0,1}) c_0  $ \\
  \vspace{2 mm}
  $c_3 = ( A_0 B_2 +A_2 B_1 +A_{0,1,2}) c_0  $\\
  \vspace{2 mm}
  $c_4 = ( B_{1,3}+ A_{0,1} B_3 + A_{0,3} B_2 + A_{2,3} B_1 + A_{0,1,2,3}) c_0  $ \\
  \vspace{2 mm}
  $c_5 = ( A_0 B_{2,4}+ A_2 B_{1,4}+ A_4 B_{1,3} + A_{0,1,2}B_4 + A_{0,1,4} B_3  + A_{0,3,4} B_2 + A_{2,3,4} B_1 + A_{0,1,2,3,4}) c_0  $  \\     
  \vspace{2 mm}
  $c_6 = (B_{1,3,5} + A_{0,1} B_{3,5} + A_{0,3} B_{2,5} + A_{0,5} B_{2,4}+ A_{2,3} B_{1,5} + A_{2,5} B_{1,4} + A_{4,5} B_{1,3} $ \\
  \vspace{2 mm}
  \hspace{0.8 cm} $ + A_{0,1,2,3} B_5 + A_{0,1,2,5}B_4+ A_{0,1,4,5} B_3  + A_{0,3,4,5} B_2 + A_{2,3,4,5} B_1 + A_{0,1,2,3,4,5}) c_0 $ \\     
  \vspace{2 mm}
  $c_7 = ( A_0 B_{2,4,6} + A_2 B_{1,4,6}+ A_4 B_{1,3,6}+ A_6 B_{1,3,5} + A_{0,1,2} B_{4,6} + A_{0,1,4} B_{3,6} + A_{0,1,6} B_{3,5} + A_{0,3,4} B_{2,6}  $ \\
  \vspace{2 mm}
  \hspace{0.8 cm} $+ A_{0,3,6} B_{2,5} + A_{0,5,6} B_{2,4} + A_{2,3,4} B_{1,6} + A_{2,3,6} B_{1,5} + A_{2,5,6} B_{1,4} + A_{4,5,6} B_{1,3}  $\\
  \vspace{2 mm}                 
  \hspace{0.8 cm} $ + A_{0,1,2,3,4}B_6 + A_{2,3,4,5,6} B_1+ A_{0,3,4,5,6} B_2+ A_{0,1,4,5,6} B_3 + A_{0,1,2,5,6}B_4 +A_{0,1,2,3,6} B_5+A_{0,1,2,3,4,5,6}) c_0 $ \\
  \vspace{2 mm}
  $c_8 = ( B_{1,3,5,7}+ A_{0,1}B_{3,5,7} + A_{0,3}B_{2,5,7} + A_{0,5}B_{2,4,7} + A_{0,7} B_{2,4,6} + A_{2,3}B_{1,5,7} + A_{2,5}B_{1,4,7} + A_{2,7} B_{1,4,6}  $ \\
  \vspace{2 mm}
  \hspace{0.8 cm} $  + A_{4,5}B_{1,3,7}+ A_{4,7} B_{1,3,6} + A_{6,7} B_{1,3,5}  + A_{0,1,2,3} B_{5,7} + A_{0,1,2,5}B_{4,7} + A_{0,1,2,7} B_{4,6}  + A_{0,1,4,5}B_{3,7}  $\\
  \vspace{2 mm}
  \hspace{0.8 cm} $+ A_{0,1,4,7} B_{3,6} + A_{0,1,6,7} B_{3,5} + A_{0,3,4,5} B_{2,7} + A_{0,3,4,7} B_{2,6} + A_{0,3,6,7} B_{2,5} + A_{0,5,6,7} B_{2,4} + A_{2,3,4,5} B_{1,7}  $\\
  \vspace{2 mm}                 
  \hspace{0.8 cm} $ + A_{2,3,4,7} B_{1,6} + A_{2,3,6,7} B_{1,5} + A_{2,5,6,7} B_{1,4}+ A_{4,5,6,7} B_{1,3}+ A_{0,1,2,3,4,5} B_{7}+ A_{0,1,2,3,4,7}B_6 $\\
   \vspace{2 mm}
   \hspace{0.8 cm} $  + A_{0,1,2,3,6,7} B_5  + A_{0,1,2,5,6,7}B_4 + A_{0,1,4,5,6,7} B_3 + A_{0,3,4,5,6,7} B_2 +A_{2,3,4,5,6,7} B_1 +A_{0,1,2,3,4,5,6,7}) c_0$\\ 
\hspace{2 mm}\large{\vdots} \hspace{5cm}\large{\vdots}\\ 
\end{tabular}
\label{eq:7}
\end{equation}
In (\ref{eq:7}) the number of individual sequence $c_n$ follows Fibonacci sequence: 1,1,2,3,5,8,13,21,34,55,$\cdots$.
The sequence $c_n$ consists of combinations $A_n$ and $B_n$ in (\ref{eq:7}). 
First observe the term inside parentheses of sequence $c_n$ which does not include any $A_n$'s: $c_n$ with even index ($c_0$, $c_2$, $c_4$,$\cdots$). 

(a) Zero term of $A_n$'s

\begin{equation}
\begin{tabular}{  l  }
  \vspace{2 mm}
  $c_0$ \\
  \vspace{2 mm}
  $c_2 = B_1 c_0  $ \\
  \vspace{2 mm}
  $c_4 = B_{1,3} c_0  $ \\
  \vspace{2 mm}
  $c_6 = B_{1,3,5}c_0 $ \\
  \vspace{2 mm}
  $c_8 = B_{1,3,5,7}c_0 $\\
  \hspace{2 mm}
  \large{\vdots}\hspace{1cm}\large{\vdots} \\ 
\end{tabular}
\label{eq:8}
\end{equation}

When a function $y(x)$, analytic at $x=0$, is expanded in a power series, we write
\begin{equation}
y(x)= \sum_{n=0}^{\infty } c_n x^{n+\lambda }= \sum_{m=0}^{\infty } y_m(x) = y_0(x)+ y_1(x)+y_2(x)+ \cdots \label{eq:9}
\end{equation}
where
\begin{equation}
y_m(x)= \sum_{l=0}^{\infty } c_l^m x^{l+\lambda }\label{eq:104}
\end{equation}
$\lambda $ is the indicial root. $y_m(x)$ is sub-power series that have sequence $c_n$ including $m$ term of $A_n$'s in (\ref{eq:7}). For example $y_0(x)$ has sequences $c_n$ including zero term of $A_n$'s in (\ref{eq:7}), $y_1(x)$ has sequences $c_n$ including one term of $A_n$'s in (\ref{eq:7}), $y_2(x)$ has sequences $c_n$ including two term of $A_n$'s in (\ref{eq:7}), etc. Substitute (\ref{eq:8}) in (\ref{eq:104}) putting $m = 0$. 
\begin{equation}
y_0(x) = c_0 \sum_{n=0}^{\infty} \left\{ \prod _{i_0=0}^{n-1}B_{2i_0+1} \right\} x^{2n+\lambda }\label{eq:10}
\end{equation}

Observe the terms inside parentheses of sequence $c_n$ which include one term of $A_n$'s in (\ref{eq:7}): $c_n$ with odd index ($c_1$, $c_3$, $c_5$,$\cdots$). 

(b) One term of $A_n$'s

\begin{equation}
\begin{tabular}{  l  }
  \vspace{2 mm}
  $c_1= A_0 c_0$ \\
  \vspace{2 mm}
  $c_3 = \left\{ A_0 \Big( \frac{B_2}{1}\Big)1+ A_2\Big(\frac{B_2}{B_2}\Big) B_1\right\} c_0  $ \\
  \vspace{2 mm}
  $c_5 = \left\{ A_0 \Big( \frac{B_{2,4}}{1}\Big)1+ A_2\Big(\frac{B_{2,4}}{B_2}\Big) B_1+ A_4\Big(\frac{B_{2,4}}{B_{2,4}}\Big) B_{1,3}\right\} c_0  $ \\
  \vspace{2 mm}
  $c_7 = \left\{ A_0 \Big( \frac{B_{2,4,6}}{1}\Big)1+ A_2\Big(\frac{B_{2,4,6}}{B_2}\Big) B_1+ A_4\Big(\frac{B_{2,4,6}}{B_{2,4}}\Big) B_{1,3}+  A_6\Big(\frac{B_{2,4,6}}{B_{2,4,6}}\Big) B_{1,3,5}\right\} c_0  $ \\
  \vspace{2 mm}
  $c_9 = \bigg\{ A_0 \Big( \frac{B_{2,4,6,8}}{1}\Big)1+ A_2\Big(\frac{B_{2,4,6,8}}{B_2}\Big) B_1+ A_4\Big(\frac{B_{2,4,6,8}}{B_{2,4}}\Big) B_{1,3}+  A_6\Big(\frac{B_{2,4,6,8}}{B_{2,4,6}}\Big) B_{1,3,5} $\\
  \vspace{2 mm}
  \hspace{0.8 cm} $+  A_8\Big(\frac{B_{2,4,6,8}}{B_{2,4,6,8}}\Big) B_{1,3,5,7}\bigg\} c_0 $\\
  $c_{11} =\bigg\{ A_0 \Big( \frac{B_{2,4,6,8,10}}{1}\Big)1+ A_2\Big(\frac{B_{2,4,6,8,10}}{B_2}\Big) B_1+ A_4\Big(\frac{B_{2,4,6,8,10}}{B_{2,4}}\Big) B_{1,3}+  A_6\Big(\frac{B_{2,4,6,8,10}}{B_{2,4,6}}\Big) B_{1,3,5}$\\
   \vspace{2 mm}
   \hspace{0.8 cm} $+  A_8\Big(\frac{B_{2,4,6,8,10}}{B_{2,4,6,8}}\Big) B_{1,3,5,7}+  A_{10}\Big(\frac{B_{2,4,6,8,10}}{B_{2,4,6,8,10}}\Big)B_{1,3,5,7,9} \bigg\} c_0 $\\
  \hspace{2 mm}
  \large{\vdots}\hspace{3cm}\large{\vdots} \\ 
\end{tabular}
\label{eq:11}
\end{equation}

(\ref{eq:11}) is simply
\begin{equation}
c_{2n+1}= c_0  \sum_{i_0=0}^{n} \left\{ A_{2i_0} \prod _{i_1=0}^{i_0-1}B_{2i_1+1} \prod _{i_2=i_0}^{n-1}B_{2i_2+2} \right\} 
\label{eq:12}
\end{equation}
Substitute (\ref{eq:12}) in (\ref{eq:104}) putting $m = 1$. 
\begin{equation}
y_1(x)= c_0 \sum_{n=0}^{\infty}\left\{ \sum_{i_0=0}^{n} \left\{ A_{2i_0} \prod _{i_1=0}^{i_0-1}B_{2i_1+1} \prod _{i_2=i_0}^{n-1}B_{2i_2+2} \right\} \right\} x^{2n+1+\lambda } \label{eq:13}
\end{equation}
Observe the terms inside parentheses of sequence $c_n$ which include two terms of $A_n$'s in (\ref{eq:7}): $c_n$ with even index ($c_2$, $c_4$, $c_6$,$\cdots$). 

(c) Two terms of $A_n$'s
\begin{equation}
\begin{tabular}{  l  }
  \vspace{2 mm}
  $c_2= A_{0,1} c_0$ \\
  \vspace{2 mm}
  $c_4 = \Bigg\{ A_0 \left\{ A_1 \left( \frac{1}{1}\right) \left(\frac{B_{1,3}}{B_1}\right)1+ A_3 \left(\frac{B_2}{1}\right) \left( \frac{B_{1,3}}{B_{1,3}}\right)1\right\} + A_2 \left\{ A_3 \left( \frac{B_2}{B_2}\right) \left( \frac{B_{1,3}}{B_{1,3}}\right) B_1 \right\} \Bigg\} c_0  $ \\
  \vspace{2 mm}
  $c_6 = \Bigg\{ A_0 \left\{ A_1 \left( \frac{1}{1}\right) \left(\frac{B_{1,3,5}}{B_1}\right)1+ A_3 \left(\frac{B_2}{1}\right) \left( \frac{B_{1,3,5}}{B_{1,3}}\right)1+ A_5 \left(\frac{B_{2,4}}{1}\right) \left( \frac{B_{1,3,5}}{B_{1,3,5}}\right)1\right\}$\\
  \vspace{2 mm}
  \hspace{0.8 cm} $+ A_2 \left\{ A_3 \left( \frac{B_2}{B_2}\right) \left( \frac{B_{1,3,5}}{B_{1,3}}\right) B_1 + A_5 \left( \frac{B_{2,4}}{B_2}\right) \left( \frac{B_{1,3,5}}{B_{1,3,5}}\right) B_1\right\} + A_4 \left\{ A_5 \left( \frac{B_{2,4}}{B_{2,4}}\right) \left( \frac{B_{1,3,5}}{B_{1,3,5}}\right) B_{1,3} \right\} \Bigg\} c_0 $ \\
\vspace{2 mm}
    $c_8 = \Bigg\{ A_0 \bigg\{ A_1 \left( \frac{1}{1}\right) \left(\frac{B_{1,3,5,7}}{B_1}\right)1+ A_3 \left(\frac{B_2}{1}\right) \left( \frac{B_{1,3,5,7}}{B_{1,3}}\right)1+ A_5 \left(\frac{B_{2,4}}{1}\right) \left( \frac{B_{1,3,5,7}}{B_{1,3,5}}\right)1 +  A_7 \left(\frac{B_{2,4,6}}{1}\right) \left( \frac{B_{1,3,5,7}}{B_{1,3,5,7}}\right)1  \bigg\}$\\
\vspace{2 mm}
  \hspace{0.8 cm} $+ A_2 \left\{ A_3 \left( \frac{B_2}{B_2}\right) \left( \frac{B_{1,3,5,7}}{B_{1,3}}\right) B_1 + A_5 \left( \frac{B_{2,4}}{B_2}\right) \left( \frac{B_{1,3,5,7}}{B_{1,3,5}}\right) B_1 + A_7 \left( \frac{B_{2,4,6}}{B_2}\right) \left( \frac{B_{1,3,5,7}}{B_{1,3,5,7}}\right) B_1\right\}$\\
\vspace{2 mm}
  \hspace{0.8 cm} $+ A_4 \left\{ A_5 \left( \frac{B_{2,4}}{B_{2,4}}\right) \left( \frac{B_{1,3,5,7}}{B_{1,3,5}}\right) B_{1,3} +  A_7 \left( \frac{B_{2,4,6}}{B_{2,4}}\right) \left( \frac{B_{1,3,5,7}}{B_{1,3,5,7}}\right) B_{1,3}\right\}$\\
\vspace{2 mm}
  \hspace{0.8 cm} $+ A_6 \left\{ A_7 \Big( \frac{B_{2,4,6}}{B_{2,4,6}}\Big) \Big( \frac{B_{1,3,5,7}}{B_{1,3,5,7}}\Big) B_{1,3,5} \right\} \Bigg\} c_0$\\
  \hspace{2 mm}
  \large{\vdots}\hspace{5cm}\large{\vdots} \\  
\end{tabular}
\label{eq:14}
\end{equation}
(\ref{eq:14}) is simply
\begin{eqnarray}
 c_{2n+2} &=& c_0\sum_{i_0=0}^{n} \left\{ A_{2i_0}\sum_{i_1=i_0}^{n} \left\{ A_{2i_1+1}{\displaystyle \prod _{i_2=0}^{i_0-1}B_{2i_2+1} \prod _{i_3=i_0}^{i_1-1}B_{2i_3+2}\prod _{i_4=i_1}^{n-1}B_{2i_4+3}}\right\}\right\}  
\nonumber\\
&=& c_0\sum_{i_0=0}^{n} \left\{ A_{2i_0} \prod _{i_1=0}^{i_0-1}B_{2i_1+1} \sum_{i_2=i_0}^{n} \left\{ A_{2i_2+1}  \prod _{i_3=i_0}^{i_2-1}B_{2i_3+2} \prod _{i_4=i_2}^{n-1}B_{2i_4+3} \right\}\right\} \label{eq:15}
\end{eqnarray}
Substitute (\ref{eq:15}) in (\ref{eq:104}) putting $m = 2$. 
\begin{equation}
 y_2(x) = c_0 \sum_{n=0}^{\infty}\left\{ \sum_{i_0=0}^{n} \left\{ A_{2i_0} \prod _{i_1=0}^{i_0-1}B_{2i_1+1} \sum_{i_2=i_0}^{n} \left\{ A_{2i_2+1}  \prod _{i_3=i_0}^{i_2-1}B_{2i_3+2} \prod _{i_4=i_2}^{n-1}B_{2i_4+3} \right\}\right\} \right\} x^{2n+2+\lambda } \label{eq:16} 
\end{equation}
Observe the terms inside parentheses of sequence $c_n$ which include three terms of $A_n$'s in (\ref{eq:7}): $c_n$ with odd index ($c_3$, $c_5$, $c_7$,$\cdots$). 

(d) Three terms of $A_n$'s
\begin{equation}
\begin{tabular}{  l  }
  \vspace{2 mm}
  $c_3= A_{0,1,2} \;c_0$ \\
  \vspace{2 mm}
  $c_5 = \Bigg\{ A_0 \Big\{ A_1 \Big[ A_2 \cdot 1 \left( \frac{1}{1}\right) \left( \frac{1}{1}\right) \left(\frac{B_{4}}{1}\right)+ A_4 \cdot 1 \left( \frac{1}{1}\right) \left(\frac{B_3}{1}\right) \left( \frac{B_{4}}{B_{4}}\right)\Big] $\\
 \vspace{2 mm}
  \hspace{0.8 cm} $+ A_3 A_4 \cdot 1 \left( \frac{B_2}{1}\right) \left( \frac{B_{3}}{B_{3}}\right) \left( \frac{B_{4}}{B_{4}}\right) \Big\} + A_2 \Big\{ A_3 \Big[ A_4 B_1 \left( \frac{B_{2}}{B_{2}}\right)\left( \frac{B_{3}}{B_{3}}\right)\left( \frac{B_{4}}{B_{4}}\right)\Big] \Big\} \Bigg\} c_0  $ \\
  \vspace{2 mm}
  $c_7 = \Bigg\{ A_0 \Big\{ A_1 \Big[ A_2 \cdot 1 \left( \frac{1}{1}\right) \left( \frac{1}{1}\right) \left(\frac{B_{4,6}}{1}\right)+ A_4 \cdot 1 \left( \frac{1}{1}\right) \left(\frac{B_3}{1}\right) \left( \frac{B_{4,6}}{B_{4}}\right)+ A_6 \cdot 1 \left( \frac{1}{1}\right) \left(\frac{B_{3,5}}{1}\right) \left( \frac{B_{4,6}}{B_{{4,6}}}\right)\Big]$\\
 \vspace{2 mm}
  \hspace{0.8 cm} $+ A_3 \Big[ A_4 \cdot 1 \left( \frac{B_2}{1}\right) \left( \frac{B_{3}}{B_{3}}\right) \left( \frac{B_{4,6}}{B_{4}}\right) + A_6 \cdot 1 \left( \frac{B_2}{1}\right) \left( \frac{B_{3,5}}{B_{3}}\right) \left( \frac{B_{4,6}}{B_{4,6}}\right)\Big]+ A_5 \Big[ A_6\cdot 1 \left( \frac{B_{2,4}}{1}\right) \left( \frac{B_{3,5}}{B_{3,5}}\right) \left( \frac{B_{4,6}}{B_{4,6}}\right) \Big] \Big\} $\\
 \vspace{2 mm}
  \hspace{0.8 cm} $+ A_2 \Big\{ A_3 \Big[ A_4 B_1 \left( \frac{B_{2}}{B_{2}}\right)\left( \frac{B_{3}}{B_{3}}\right)\left( \frac{B_{4,6}}{B_{4}}\right)+ A_6 B_1 \left( \frac{B_{2}}{B_{2}}\right)\left( \frac{B_{3,5}}{B_{3}}\right)\left( \frac{B_{4,6}}{B_{4,6}}\right)\Big] $\\
\vspace{2 mm}
  \hspace{0.8 cm}$ + A_5 \Big[ A_6 B_1 \left( \frac{B_{2,4}}{B_{2}}\right) \left( \frac{B_{3,5}}{B_{3,5}}\right) \left( \frac{B_{4,6}}{B_{4,6}}\right) \Big] \Big\}+ A_4 \Big\{ A_5 \Big[ A_6 B_{1,3} \left( \frac{B_{2,4}}{B_{2,4}}\right) \left( \frac{B_{3,5}}{B_{3,5}}\right) \left( \frac{B_{4,6}}{B_{4,6}}\right) \Big]\Big\} \Bigg\} c_0 $ \\
 \vspace{2 mm}
   $c_9 = \Bigg\{ A_0 \Big\{ A_1 \Big[ A_2 \cdot 1 \left( \frac{1}{1}\right) \left( \frac{1}{1}\right) \left(\frac{B_{4,6,8}}{1}\right)+ A_4 \cdot 1 \left( \frac{1}{1}\right) \left(\frac{B_3}{1}\right) \left( \frac{B_{4,6,8}}{B_{4}}\right)+ A_6 \cdot 1 \left( \frac{1}{1}\right) \left(\frac{B_{3,5}}{1}\right) \left( \frac{B_{4,6,8}}{B_{{4,6}}}\right) + A_8 \cdot 1 \left( \frac{1}{1}\right) \left(\frac{B_{3,5,7}}{1}\right) \left( \frac{B_{4,6,8}}{B_{{4,6,8}}}\right)\Big]$\\
\vspace{2 mm}
\hspace{0.8 cm} $+ A_3 \Big[ A_4 \cdot 1 \left( \frac{B_2}{1}\right) \left( \frac{B_{3}}{B_{3}}\right) \left( \frac{B_{4,6,8}}{B_{4}}\right) + A_6 \cdot 1 \left( \frac{B_2}{1}\right) \left( \frac{B_{3,5}}{B_{3}}\right) \left( \frac{B_{4,6,8}}{B_{4,6}}\right)+ A_8 \cdot 1 \left( \frac{B_2}{1}\right) \left( \frac{B_{3,5,7}}{B_{3}}\right) \left( \frac{B_{4,6,8}}{B_{4,6,8}}\right) \Big]$\\
  \vspace{2 mm}
  \hspace{0.8 cm} $+ A_5 \Big[ A_6\cdot 1 \left( \frac{B_{2,4}}{1}\right) \left( \frac{B_{3,5}}{B_{3,5}}\right) \left( \frac{B_{4,6,8}}{B_{4,6}}\right)+ A_8\cdot 1 \left( \frac{B_{2,4}}{1}\right) \left( \frac{B_{3,5,7}}{B_{3,5}}\right) \left( \frac{B_{4,6,8}}{B_{4,6,8}}\right)\Big] + A_7\Big[ A_8\cdot 1 \left( \frac{B_{2,4,6}}{1}\right) \left( \frac{B_{3,5,7}}{B_{3,5,7}}\right) \left( \frac{B_{4,6,8}}{B_{4,6,8}}\right) \Big] \Big\} $\\
 \vspace{2 mm}
  \hspace{0.8 cm} $+ A_2 \Big\{ A_3 \Big[ A_4 B_1 \left( \frac{B_{2}}{B_{2}}\right)\left( \frac{B_{3}}{B_{3}}\right)\left( \frac{B_{4,6,8}}{B_{4}}\right)+ A_6 B_1 \left( \frac{B_{2}}{B_{2}}\right)\left( \frac{B_{3,5}}{B_{3}}\right)\left( \frac{B_{4,6,8}}{B_{4,6}}\right) + A_8 B_1 \left( \frac{B_{2}}{B_{2}}\right)\left( \frac{B_{3,5,7}}{B_{3}}\right)\left( \frac{B_{4,6,8}}{B_{4,6,8}}\right)\Big]$\\
  \vspace{2 mm}
  \hspace{0.8 cm} $+ A_5 \Big[ A_6 B_1 \left( \frac{B_{2,4}}{B_{2}}\right) \left( \frac{B_{3,5}}{B_{3,5}}\right) \left( \frac{B_{4,6,8}}{B_{4,6}}\right) + A_8 B_1 \left( \frac{B_{2,4}}{B_{2}}\right) \left( \frac{B_{3,5,7}}{B_{3,5}}\right) \left( \frac{B_{4,6,8}}{B_{4,6,8}}\right)\Big] + A_7 \Big[ A_8 B_1 \left( \frac{B_{2,4,6}}{B_{2}}\right) \left( \frac{B_{3,5,7}}{B_{3,5,7}}\right) \left( \frac{B_{4,6,8}}{B_{4,6,8}}\right) \Big] \Big\}$\\
\vspace{2 mm}
  \hspace{0.8 cm} $+ A_4 \Big\{ A_5 \Big[ A_6 B_{1,3} \left( \frac{B_{2,4}}{B_{2,4}}\right) \left( \frac{B_{3,5}}{B_{3,5}}\right) \left( \frac{B_{4,6,8}}{B_{4,6}}\right) + A_8 B_{1,3} \left( \frac{B_{2,4}}{B_{2,4}}\right) \left( \frac{B_{3,5,7}}{B_{3,5}}\right) \left( \frac{B_{4,6,8}}{B_{4,6,8}}\right)\Big] $\\
\vspace{2 mm}
 \hspace{0.8 cm} $+ A_7 \Big[ A_8 B_{1,3} \left( \frac{B_{2,4,6}}{B_{2,4}}\right) \left( \frac{B_{3,5,7}}{B_{3,5,7}}\right) \left( \frac{B_{4,6,8}}{B_{4,6,8}}\right) \Big] \Big\}+ A_6 \Big\{ A_7 \Big[ A_8 B_{1,3,5}  \left( \frac{B_{2,4,6}}{B_{2,4,6}}\right) \left( \frac{B_{3,5,7}}{B_{3,5,7}}\right) \left( \frac{B_{4,6,8}}{B_{4,6,8}}\right)\Big] \Big\} \Bigg\} c_0 $ \\
  \hspace{2 mm}
  \large{\vdots}\hspace{6cm}\large{\vdots} \\ \\  
\end{tabular}
\label{eq:17}
\end{equation}
(\ref{eq:17}) is simply
\begin{eqnarray}
 c_{2n+3} &=& c_0 \sum_{i_0=0}^{n} \left\{ A_{2i_0}\sum_{i_1=i_0}^{n} \left\{ A_{2i_1+1}\sum_{i_2=i_1}^{n} \left\{ A_{2i_2+2} \prod _{i_3=0}^{i_0-1}B_{2i_3+1} \prod _{i_4=i_0}^{i_1-1}B_{2i_4+2} \prod _{i_5=i_1}^{i_2-1}B_{2i_5+3}\prod _{i_6=i_2}^{n-1}B_{2i_6+4} \right\} \right\} \right\} \nonumber\\
&=& c_0 \sum_{i_0=0}^{n} \left\{ A_{2i_0}\prod _{i_1=0}^{i_0-1}B_{2i_1+1} \sum_{i_2=i_0}^{n} \left\{ A_{2i_2+1} \prod _{i_3=i_0}^{i_2-1}B_{2i_3+2}  \sum_{i_4=i_2}^{n} \left\{ A_{2i_4 +2} \prod _{i_5=i_2}^{i_4-1}B_{2i_5+3}\prod _{i_6=i_4}^{n-1}B_{2i_6+4} \right\} \right\} \right\} \hspace{1cm}\label{eq:18} 
\end{eqnarray}
Substitute (\ref{eq:18}) in (\ref{eq:104}) putting $m = 3$. 
\begin{eqnarray}
 y_3(x) &=& c_0 \sum_{n=0}^{\infty }\left\{ \sum_{i_0=0}^{n} \left\{ A_{2i_0}\prod _{i_1=0}^{i_0-1}B_{2i_1+1} \sum_{i_2=i_0}^{n} \left\{ A_{2i_2+1} \prod _{i_3=i_0}^{i_2-1}B_{2i_3+2}\right.\right.\right. \nonumber\\
&&\times \left.\left.\left. \sum_{i_4=i_2}^{n} \left\{ A_{2i_4 +2} \prod _{i_5=i_2}^{i_4-1}B_{2i_5+3}\prod _{i_6=i_4}^{n-1}B_{2i_6+4} \right\} \right\} \right\}\right\} x^{2n+3+\lambda } \label{eq:19}
\end{eqnarray}
By repeating this process for all higher terms of $A$'s, we obtain every $y_m(x)$ terms where $m \geq 4$. Substitute (\ref{eq:10}), (\ref{eq:13}), (\ref{eq:16}), (\ref{eq:19}) and including all $y_m(x)$ terms where $m \geq 4$ into (\ref{eq:9}). 
\begin{thm}
The general expression of $y(x)$ for infinite series is
\begin{eqnarray}
 y(x) &=& y_0(x)+ y_1(x)+ y_2(x)+y_3(x)+\cdots \nonumber\\
&=& c_0 \left\{ \sum_{n=0}^{\infty} \left\{ \prod _{i_0=0}^{n-1}B_{2i_0+1} \right\} x^{2n+\lambda} + \sum_{n=0}^{\infty}\left\{ \sum_{i_0=0}^{n} \left\{ A_{2i_0} \prod _{i_1=0}^{i_0-1}B_{2i_1+1} \prod _{i_2=i_0}^{n-1}B_{2i_2+2} \right\} \right\} x^{2n+1+\lambda} \right. \nonumber\\
&&+ \sum_{N=2}^{\infty }\left\{ \sum_{n=0}^{\infty } \left\{ \sum_{i_0=0}^{n} \left\{A_{2i_0}\prod _{i_1=0}^{i_0-1} B_{2i_1+1} 
\prod _{k=1}^{N-1} \left( \sum_{i_{2k}= i_{2(k-1)}}^{n} A_{2i_{2k}+k}\prod _{i_{2k+1}=i_{2(k-1)}}^{i_{2k}-1}B_{2i_{2k+1}+k+1}\right) \right.\right.\right. \nonumber\\
&& \times \left.\left.\left.\left. \prod _{i_{2N}=i_{2(N-1)}}^{n-1} B_{2i_{2N}+N+1} \right\}\right\}\right\}x^{2n+N+\lambda} \right\} \label{eq:20}
\end{eqnarray}
\end{thm}
\section{\label{sec:level4}Polynomial which makes $B_n$ term terminated}
Now let's investigate the polynomial case of $y(x)$. Assume that $B_n$ is terminated at certain value of n. Then each $y_i(x)$ where $i=0,1,2,\cdots$ will be polynomial. Examples of these are Heun's equation, GCH function\cite{Ch2012}, Lame function, etc. First $B_{2k+1}$ is terminated at certain value of k. I choose eigenvalue $\beta_0$ which $B_{2k+1}$ is terminated where $\beta _0 =0,1,2,\cdots$. $B_{2k+2}$ is terminated at certain value of k. I choose eigenvalue $\beta _1$ which $B_{2k+2}$ is terminated where $\beta _1 =0,1,2,\cdots$. Also $B_{2k+3}$ is terminated at certain value of k. I choose eigenvalue $\beta _2$ which $B_{2k+3}$ is terminated where $\beta _2 =0,1,2,\cdots$. By repeating this process I obtain
\begin{equation}
B_{2\beta _i+(i+1)}=0 \hspace{1cm} \mathrm{where}\;i,\beta _i =0,1,2,\cdots
\label{eq:21}
\end{equation}
In general, the two term recurrence relation for polynomial has only one eigenvalue: for example, the Laguerre function, confluent hypergeometric function, Legendre function, etc. But three term recurrence relation has infinite eigenvalues which are $\beta _i$, where $i,\beta _i =0,1,2,\cdots$. 

First observe the term in sequence $c_n$ which does not include any $A_n$'s in (\ref{eq:8}): $c_n$ with even index ($c_0$, $c_2$, $c_4$,$\cdots$). 

(a) As $\beta _0$=0, then $B_1$=0 in (\ref{eq:8}).
\begin{equation}
\begin{tabular}{  l  }
  \vspace{2 mm}
  $c_0$ \\ \hspace{1.6cm}
\end{tabular}
\label{eq:22}
\end{equation}
(b) As $\beta _0$=1, then $B_3$=0 in (\ref{eq:8}).
\begin{equation}
\begin{tabular}{  l  }
  \vspace{2 mm}
  $c_0$ \\
  \vspace{2 mm}
  $c_2 = B_1 c_0  $ \\
\end{tabular}
\label{eq:23}
\end{equation}
(c) As $\beta _0$=2, then $B_5$=0 in (\ref{eq:8}).
\begin{equation}
\begin{tabular}{  l  }
  \vspace{2 mm}
  $c_0$ \\
  \vspace{2 mm}
  $c_2 = B_1 c_0  $ \\
  \vspace{2 mm}
  $c_4 = B_{1,3} c_0  $ \\
\end{tabular}
\label{eq:24}
\end{equation}
Substitute (\ref{eq:22}),(\ref{eq:23}) and (\ref{eq:24}) in (\ref{eq:104}) putting $m = 0$.
\begin{equation}
y_0(x)= c_0 \sum_{n=0}^{\beta _0} \left\{ \prod _{i_1=0}^{n-1}B_{2i_1+1} \right\} x^{2n+\lambda }
\label{eq:25}
\end{equation} 
Observe the terms inside curly brackets of sequence $c_n$ which include one term of $A_n$'s in (\ref{eq:11}): $c_n$ with odd index ($c_1$, $c_3$, $c_5$,$\cdots$). 

(a) As $\beta _0$=0, then $B_1$=0 in (\ref{eq:11}).
\begin{equation}
\begin{tabular}{  l  }
  \vspace{2 mm}
  $c_1= A_0 c_0$ \\
  \vspace{2 mm}
  $c_3 = A_0 B_2 c_0  $ \\
  \vspace{2 mm}
  $c_5 = A_0 B_{2,4}c_0 $ \\
  \vspace{2 mm}
  $c_7 = A_0 B_{2,4,6}c_0  $ \\
  \vspace{2 mm}
  $c_9 = A_0 B_{2,4,6,8}c_0  $ \\
  \hspace{2 mm}
  \large{\vdots}\hspace{1cm}\large{\vdots} \\ 
\end{tabular}
\label{eq:26}
\end{equation}
As i=1 in (\ref{eq:21}),
\begin{equation}
B_{2\beta _1 +2}=0 \hspace{1cm} \mathrm{where}\; \beta _1 =0,1,2,\cdots
\label{eq:27}
\end{equation}
Substitute (\ref{eq:26}) in (\ref{eq:104}) putting $m = 1$ by using (\ref{eq:27}).
\begin{equation}
y_1^0(x)= c_0 A_0 \sum_{n=0}^{\beta _1} \left\{ \prod _{i_1=0}^{n-1}B_{2i_1+2} \right\} x^{2n+1+\lambda }  
\label{eq:28}
\end{equation}
In (\ref{eq:28}) $y_1^0(x)$ is sub-power series, having sequences $c_n$ including one term of $A_n$'s in (\ref{eq:7}) as $\beta _0$=0, for the polynomial case in which makes $B_n$ term terminated.

(b) As $\beta _0$=1, then $B_3$=0 in (\ref{eq:11}).
\begin{equation}
\begin{tabular}{  l  }
  \vspace{2 mm}
  $c_1= A_0 c_0$ \\
  \vspace{2 mm}
  $c_3 = \{A_0 B_2\cdot 1+ A_2\cdot 1 B_1 \} c_0  $ \\
  \vspace{2 mm}
  $c_5 = \{A_0 B_{2,4}\cdot 1+ A_2 B_4 B_1\} c_0 $ \\
  \vspace{2 mm}
  $c_7 = \{A_0 B_{2,4,6}\cdot 1+ A_2 B_{4,6} B_1 \} c_0  $ \\
  \vspace{2 mm}
  $c_9 = \{A_0 B_{2,4,6,8}\cdot 1+ A_2 B_{4,6,8} B_1\} c_0  $ \\
  \hspace{2 mm}
  \large{\vdots}\hspace{2cm}\large{\vdots} \\ 
\end{tabular}
\label{eq:29}
\end{equation}
The first term in curly brackets of sequence $c_n$ in (\ref{eq:29}) is same as (\ref{eq:26}). Then, its solution is equal to (\ref{eq:28}). Substitute (\ref{eq:27}) into the second term in curly brackets of sequence $c_n$ in (\ref{eq:29}). Its power series expansion including the first and second terms in curly brackets of sequence $c_n$ in (\ref{eq:29}), analytic at $x=0$, is
\begin{equation}
y_1^1(x)= c_0 \left\{ A_0 \sum_{n=0}^{\beta _1} \left\{ \prod _{i_1=0}^{n-1}B_{2i_1+2} \right\} + A_2 B_1 \sum_{n=1}^{\beta _1} \left\{ \prod _{i_1=1}^{n-1}B_{2i_1+2} \right\}\right\} x^{2n+1+\lambda }  
\label{eq:30}
\end{equation}
In (\ref{eq:30}) $y_1^1(x)$ is sub-power series, having sequences $c_n$ including one term of $A_n$'s in (\ref{eq:7}), as $\beta _0$=1 for the polynomial case in which makes $B_n$ term terminated.

(c) As $\beta _0$=2, then $B_5$=0 in (\ref{eq:11}).
\begin{equation}
\begin{tabular}{  l  }
  \vspace{2 mm}
  $c_1= A_0 c_0$ \\
  \vspace{2 mm}
  $c_3 = \{A_0 B_2\cdot 1+ A_2\cdot 1 B_1 \} c_0  $ \\
  \vspace{2 mm}
  $c_5 = \{A_0 B_{2,4}\cdot 1+ A_2 B_4 B_1+ A_4\cdot 1 B_{1,3}\} c_0 $ \\
  \vspace{2 mm}
  $c_7 = \{A_0 B_{2,4,6}\cdot 1+ A_2 B_{4,6} B_1 + A_4 B_6 B_{1,3}\} c_0  $ \\
  \vspace{2 mm}
  $c_9 = \{A_0 B_{2,4,6,8}\cdot 1+ A_2 B_{4,6,8} B_1+ A_4 B_{6,8} B_{1,3}\} c_0  $ \\
  \vspace{2 mm}
  $c_{11} = \{A_0 B_{2,4,6,8}\cdot 1+ A_2 B_{4,6,8} B_1+ A_4 B_{6,8,10} B_{1,3}\} c_0  $ \\
  \hspace{2 mm}
  \large{\vdots}\hspace{4cm}\large{\vdots} \\ 
\end{tabular}
\label{eq:31}
\end{equation}
The first and second term in curly brackets of sequence $c_n$ in (\ref{eq:31}) is same as (\ref{eq:29}). Then its power series expansion is same as (\ref{eq:30}). Substitute (\ref{eq:27}) into the third term in curly brackets of sequence $c_n$ in (\ref{eq:31}). Its power series expansion including the first, second and third terms in curly brackets of sequence $c_n$ in (\ref{eq:31}), analytic at $x=0$, is
\begin{eqnarray}
y_1^2(x)&=& c_0 \left\{ A_0 \sum_{n=0}^{\beta _1} \left\{ \prod _{i_1=0}^{n-1}B_{2i_1+2} \right\} + A_2 B_1 \sum_{n=1}^{\beta _1} \left\{ \prod _{i_1=1}^{n-1}B_{2i_1+2} \right\}\right.\nonumber\\
&&+\left. A_4 B_{1,3} \sum_{n=2}^{\beta _1} \left\{ \prod _{i_1=2}^{n-1}B_{2i_1+2} \right\}\right\} x^{2n+1+\lambda } 
 \label{eq:32}
\end{eqnarray}
In (\ref{eq:32}) $y_1^2(x)$ is sub-power series, having sequence $c_n$ including one term of $A_n$'s in (\ref{eq:7}) as $\beta _0$=2, for the polynomial case in which makes $B_n$ term terminated.
By repeating this process for all $\beta _0 =3,4,5,\cdots$, I obtain every $y_1^j(x)$ terms where $j \geq 3$. 
According to (\ref{eq:28}), (\ref{eq:30}), (\ref{eq:32}) and every $y_1^j(x)$ where $j \geq 3$, the general expression of $y_1(x)$ for all $\beta _0$ is 
\begin{equation}
y_1(x)= c_0 \sum_{i_0=0}^{\beta _0}\left\{ A_{2i_0} \prod _{i_1=0}^{i_0-1}B_{2i_1+1}  \sum_{i_2=i_0}^{\beta _1} \left\{ \prod _{i_3=i_0}^{i_2-1}B_{2i_3+2} \right\}\right\} x^{2i_2+1+\lambda } \;\;\mbox{where}\; \beta _0 \leq \beta _1
 \label{eq:33} 
\end{equation}

Observe the terms of sequence $c_n$ which include two terms of $A_n$'s in (\ref{eq:14}): $c_n$ with even index ($c_2$, $c_4$, $c_6$,$\cdots$). 

(a) As $\beta _0$=0, then $B_1$=0 in (\ref{eq:14}).
\begin{equation}
\begin{tabular}{  l  }
  \vspace{2 mm}
  $c_2= A_{0,1} c_0$ \\
  \vspace{2 mm}
  $c_4 = A_0\{A_1\cdot 1 \cdot B_3\cdot 1+ A_3 B_2\cdot 1 \cdot 1 \} c_0  $ \\
  \vspace{2 mm}
  $c_6 = A_0 \{A_1\cdot 1 \cdot B_{3,5}\cdot 1+ A_3 B_2 B_5\cdot 1+ A_5 B_{2,4}\cdot 1\cdot 1\} c_0 $ \\
  \vspace{2 mm}
  $c_8 = A_0 \{A_1\cdot 1 \cdot B_{3,5,7}\cdot 1+ A_3 B_2 B_{5,7}\cdot 1+ A_5 B_{2,4}B_7\cdot 1+A_7 B_{2,4,6}\cdot 1\cdot 1\} c_0  $ \\
  \vspace{2 mm}
  $c_{10} = A_0 \{A_1\cdot 1 \cdot B_{3,5,7,9}\cdot 1+ A_3 B_2 B_{5,7,9}\cdot 1+ A_5 B_{2,4}B_{7,9}\cdot 1  +A_7 B_{2,4,6}B_{9}\cdot 1$\\
\vspace{2 mm}
 \hspace{1 cm}$+ A_9 B_{2,4,6,8}\cdot 1\cdot 1\} c_0$ \\
  \large{\vdots}\hspace{5cm}\large{\vdots} \\ 
\end{tabular}
\label{eq:34}
\end{equation}
(i) As $\beta _1$=0, then $B_2$=0 in (\ref{eq:34}).
\begin{equation}
\begin{tabular}{  l  }
  \vspace{2 mm}
  $c_2= A_{0,1} c_0$ \\
  \vspace{2 mm}
  $c_4 = A_0 A_1\cdot 1 \cdot B_3\cdot 1 c_0  $ \\
  \vspace{2 mm}
  $c_6 = A_0 A_1\cdot 1 \cdot B_{3,5}\cdot 1 c_0 $ \\
  \vspace{2 mm}
  $c_8 = A_0 A_1\cdot 1 \cdot B_{3,5,7}\cdot 1 c_0  $ \\
  \vspace{2 mm}
  $c_{10} = A_0 A_1\cdot 1 \cdot B_{3,5,7,9}\cdot 1 c_0  $ \\
  \hspace{2 mm}
  \large{\vdots}\hspace{2cm}\large{\vdots} \\ 
\end{tabular}
\label{eq:35}
\end{equation}
As i=2 in (\ref{eq:21}),
\begin{equation}
B_{2\beta _2 +3}=0 \hspace{1cm} \mathrm{where}\; \beta _2 =0,1,2,\cdots
\label{eq:36}
\end{equation}
Substitute (\ref{eq:35}) in (\ref{eq:104}) putting $m = 2$ by using (\ref{eq:36}).
\begin{equation}
y_2^{0,0}(x)= c_0 A_0 A_1 \sum_{n=0}^{\beta _2} \left\{ \prod _{i_1=0}^{n-1}B_{2i_1+3} \right\} x^{2n+2+\lambda }  
\label{eq:37}
\end{equation} 
In (\ref{eq:37}) $y_2^{0,0}(x)$ is sub-power series, having sequences $c_n$ including two term of $A_n$'s in (\ref{eq:7}) as $\beta _0$=0 and $\beta _1$=0, for the polynomial case in which makes $B_n$ term terminated.

(ii) As $\beta _1$=1, then $B_4$=0 in (\ref{eq:34}).
\begin{equation}
\begin{tabular}{  l  }
  \vspace{2 mm}
  $c_2= A_{0,1} c_0$ \\
  \vspace{2 mm}
  $c_4 = A_0\{A_1\cdot 1 \cdot B_3\cdot 1+ A_3 B_2\cdot 1 \cdot 1 \} c_0  $ \\
  \vspace{2 mm}
  $c_6 = A_0 \{A_1\cdot 1 \cdot B_{3,5}\cdot 1+ A_3 B_2 B_5\cdot 1\} c_0 $ \\
  \vspace{2 mm}
  $c_8 = A_0 \{A_1\cdot 1 \cdot B_{3,5,7}\cdot 1+ A_3 B_2 B_{5,7}\cdot 1\} c_0  $ \\
  \vspace{2 mm}
  $c_{10} = A_0 \{A_1\cdot 1 \cdot B_{3,5,7,9}\cdot 1+ A_3 B_2 B_{5,7,9}\cdot 1\} c_0  $ \\
  \hspace{2 mm}
  \large{\vdots}\hspace{3cm}\large{\vdots} \\ 
\end{tabular}
\label{eq:38}
\end{equation}
The first term in curly brackets of sequence $c_n$ in (\ref{eq:38}) is same as (\ref{eq:35}). Then its power series expansion is equal to (\ref{eq:37}). Substitute (\ref{eq:36}) into the second term in curly brackets of sequence $c_n$ in (\ref{eq:38}). Its power series expansion including the first and second terms in curly brackets of sequence $c_n$, analytic at $x=0$, is
\begin{equation}
y_2^{0,1}(x)= c_0 A_0 \left\{ A_1 \sum_{n=0}^{\beta _2} \left\{ \prod _{i_1=0}^{n-1}B_{2i_1+3} \right\} + A_3 B_2 \sum_{n=1}^{\beta _2} \left\{ \prod _{i_1=1}^{n-1}B_{2i_1+3} \right\} \right\} x^{2n+2+\lambda }
\label{eq:39}
\end{equation}
In (\ref{eq:39}) $y_2^{0,1}(x)$ is sub-power series, having sequences $c_n$ including two term of $A_n$'s in (\ref{eq:7}) as $\beta _0$=0 and $\beta _1$=1, for the polynomial case in which makes $B_n$ term terminated.

(iii) As $\beta _1$=2, then $B_6$=0 in (\ref{eq:34}).
\begin{equation}
\begin{tabular}{  l  }
  \vspace{2 mm}
  $c_2= A_{0,1} c_0$ \\
  \vspace{2 mm}
  $c_4 = A_0\{A_1\cdot 1 \cdot B_3\cdot 1+ A_3 B_2\cdot 1 \cdot 1 \} c_0  $ \\
  \vspace{2 mm}
  $c_6 = A_0 \{A_1\cdot 1 \cdot B_{3,5}\cdot 1+ A_3 B_2 B_5\cdot 1+ A_5 B_{2,4}\cdot 1\cdot 1\} c_0 $ \\
  \vspace{2 mm}
  $c_8 = A_0 \{A_1\cdot 1 \cdot B_{3,5,7}\cdot 1+ A_3 B_2 B_{5,7}\cdot 1+ A_5 B_{2,4}B_7\cdot 1\} c_0  $ \\
  \vspace{2 mm}
  $c_{10} = A_0 \{A_1\cdot 1 \cdot B_{3,5,7,9}\cdot 1+ A_3 B_2 B_{5,7,9}\cdot 1+ A_5 B_{2,4}B_{7,9}\cdot 1\} c_0  $ \\
  \hspace{2 mm}
  \large{\vdots}\hspace{4cm}\large{\vdots} \\ 
\end{tabular}
\label{eq:40}
\end{equation}
The first and second term in curly brackets of sequence $c_n$ in (\ref{eq:40}) is same as (\ref{eq:38}). Then its power series expansion is same as (\ref{eq:39}). Substitute (\ref{eq:36}) into the third term in curly brackets of sequence $c_n$ in (\ref{eq:40}). Its power series expansion including the first, second and third terms in curly brackets of sequence $c_n$ in (\ref{eq:40}), analytic at $x=0$, is
\begin{eqnarray}
y_2^{0,2}(x) &=& c_0 A_0 \left\{ A_1 \sum_{n=0}^{\beta _2} \left\{ \prod _{i_1=0}^{n-1}B_{2i_1+3} \right\} + A_3 B_2 \sum_{n=1}^{\beta _2} \left\{ \prod _{i_1=1}^{n-1}B_{2i_1+3} \right\}\right.\nonumber\\
&&+\left. A_5 B_{2,4} \sum_{n=2}^{\beta _2} \left\{ \prod _{i_1=2}^{n-1}B_{2i_1+3} \right\}\right\} x^{2n+2+\lambda }
\label{eq:41}
\end{eqnarray}
In (\ref{eq:41}) $y_2^{0,2}(x)$ is sub-power series, having sequences $c_n$ including two term of $A_n$'s in (\ref{eq:7}) as $\beta _0$=0 and $\beta _1$=2, for the polynomial case in which makes $B_n$ term terminated.
By repeating this process for all $\beta _1 =3,4,5,\cdots$, we obtain every $y_2^{0,j}(x)$ terms where $j \geq 3$. 
According to (\ref{eq:37}), (\ref{eq:39}), (\ref{eq:41}) and every $y_2^{0,j}(x)$ where $j \geq 3$, the general expression of $y_2^0(x)$ for the case of $\beta _0=0$ replacing the index n by $i_0$ is 
\begin{equation}
y_2^0(x)= c_0 A_0 \sum_{i_0=0}^{\beta _1}\left\{ A_{2i_0+1} \prod _{i_1=0}^{i_0-1}B_{2i_1+2}  \sum_{i_2=i_0}^{\beta _2} \left\{ \prod _{i_3=i_0}^{i_2-1}B_{2i_3+3} \right\}\right\} x^{2i_2+2+\lambda }  
\label{eq:42}
\end{equation}
In (\ref{eq:42}) $y_2^0(x)$ is sub-power series, having sequences $c_n$ including two term of $A_n$'s in (\ref{eq:7}) as $\beta _0$=0, for the polynomial case in which makes $B_n$ term terminated.
\vspace{2 mm}

(b) As $\beta _0$=1, then $B_3$=0 in (\ref{eq:14}).
\begin{equation}
\begin{tabular}{  l  }
  \vspace{2 mm}
  $c_2= A_{0,1} c_0$ \\
  \vspace{2 mm}
  $c_4 = \Big\{ A_0\Big[ A_1\cdot 1 \cdot B_3\cdot 1+ A_3 B_2\cdot 1 \cdot 1\Big]+ A_2\Big[ A_3\cdot 1\cdot 1\cdot B_1\Big] \Big\} c_0  $ \\
  \vspace{2 mm}
  $c_6 = \Big\{ A_0 \Big[ A_1\cdot 1 \cdot B_{3,5}\cdot 1+ A_3 B_2 B_5\cdot 1+ A_5 B_{2,4}\cdot 1\cdot 1 \Big]$\\
\vspace{2 mm}
  \hspace{0.8 cm} $+ A_2\Big[ A_3\cdot 1\cdot B_5 B_1+ A_5 B_4 \cdot 1\cdot B_1\Big] \Big\} c_0$ \\
  \vspace{2 mm}
  $c_8 = \Big\{ A_0 \Big[ A_1\cdot 1 \cdot B_{3,5,7}\cdot 1+ A_3 B_2 B_{5,7}\cdot 1+ A_5 B_{2,4}B_7\cdot 1+A_7 B_{2,4,6}\cdot 1\cdot 1 \Big] $ \\
 \vspace{2 mm}
  \hspace{0.8 cm} $+A_2 \Big[ A_3\cdot 1\cdot B_{5,7} B_1+ A_5 B_4 B_7 B_1+ A_7 B_{4,6} \cdot 1\cdot B_1\Big] \Big\} c_0 $\\
  \vspace{2 mm}
  $c_{10} = \Big\{ A_0 \Big[ A_1\cdot 1 \cdot B_{3,5,7,9}\cdot 1+ A_3 B_2 B_{5,7,9}\cdot 1+ A_5 B_{2,4}B_{7,9}\cdot 1 $\\
\vspace{2 mm}
  \hspace{0.8 cm} $+A_7 B_{2,4,6}B_{9}\cdot 1+ A_9 B_{2,4,6,8}\cdot 1\cdot 1 \Big]$ \\
  \vspace{2 mm}
  \hspace{0.8 cm} $+A_2 \Big[ A_3\cdot 1\cdot B_{5,7,9} B_1+ A_5 B_4 B_{7,9} B_1+ A_7 B_{4,6} B_9 B_1+ A_9 B_{4,6,8}\cdot 1\cdot B_1\Big] \Big\} c_0 $\\ 
  \hspace{2 mm}
  \large{\vdots}\hspace{7cm}\large{\vdots} \\ 
\end{tabular}
\label{eq:43}
\end{equation}
The first square brackets including $A_0$ inside curly brackets in sequence $c_n$ in (\ref{eq:43}) is same as (\ref{eq:34}). Then its power series expansion is same as (\ref{eq:42}). Observe the second square brackets including $A_2$ inside curly brackets in sequence $c_n$ in (\ref{eq:43}).

(i) As $\beta _1$=1, then $B_4$=0 in the second square brackets including $A_2$ inside curly brackets in sequence $c_n$ in (\ref{eq:43}).
\begin{equation}
\begin{tabular}{  l  }
  \vspace{2 mm}
  $c_4= A_{2} B_1 \Big\{ A_3\cdot 1\cdot 1 \Big\} c_0$ \\
  \vspace{2 mm}
  $c_6 =A_{2} B_1 \Big\{ A_3\cdot 1\cdot B_5 \Big\}c_0 $ \\
  \vspace{2 mm}
  $c_8 = A_{2} B_1 \Big\{ A_3\cdot 1\cdot B_{5,7} \Big\} c_0  $ \\
  \vspace{2 mm}
  $c_{10} = A_{2} B_1 \Big\{ A_3\cdot 1\cdot B_{5,7,9} \Big\} c_0  $ \\
  \hspace{2 mm}
  \large{\vdots}\hspace{2cm}\large{\vdots} \\ 
\end{tabular}
\label{eq:44}
\end{equation}
Its power series expansion of (\ref{eq:44}) by using (\ref{eq:36}), analytic at $x=0$, is
\begin{equation}
y_2^{1,1}(x)= c_0 A_2 B_1 A_3 \sum_{n=1}^{\beta _2}\left\{ \prod _{i_1=1}^{n-1}B_{2i_1+3} \right\} x^{2n+2+\lambda } 
\label{eq:45}
\end{equation}
In (\ref{eq:45}) $y_2^{1,1}(x)$ is sub-power series, for the second square brackets inside curly brackets in sequence $c_n$ including two term of $A_n$'s in (\ref{eq:43}) as $\beta _0$=1 and $\beta _1$=1, for the polynomial case in which makes $B_n$ term terminated.

(ii) As $\beta _1$=2, then $B_6$=0 in second square brackets inside curly brackets in sequence $c_n$ including $A_2$ in (\ref{eq:43}).
\begin{equation}
\begin{tabular}{  l  }
  \vspace{2 mm}
  $c_4= A_{2} B_1 \Big\{ A_3\cdot 1\cdot 1 \Big\} c_0$ \\
  \vspace{2 mm}
  $c_6 =A_{2} B_1 \Big\{ A_3\cdot 1\cdot B_5 + A_5 B_4 \cdot 1 \Big\}c_0 $ \\
  \vspace{2 mm}
  $c_8 = A_{2} B_1 \Big\{ A_3\cdot 1\cdot B_{5,7} + A_5 B_4 B_7 \Big\} c_0  $ \\
  \vspace{2 mm}
  $c_{10} = A_{2} B_1 \Big\{ A_3\cdot 1\cdot B_{5,7,9} + A_5 B_4 B_{7,9} \Big\} c_0  $ \\
  \hspace{2 mm}
  \large{\vdots}\hspace{3cm}\large{\vdots} \\ 
\end{tabular}
\label{eq:46}
\end{equation}
The first term in curly brackets of sequence $c_n$ in (\ref{eq:46}) is same as (\ref{eq:44}). Then its power series expansion is same as (\ref{eq:45}). Substitute (\ref{eq:36}) into the second term in curly brackets of sequence $c_n$ in (\ref{eq:46}). Its power series expansion including the first and second terms in curly brackets of sequence $c_n$ in (\ref{eq:46}), analytic at $x=0$, is
\begin{equation}
y_2^{1,2}(x) = c_0 A_2 B_1 \left\{ A_3 \sum_{n=1}^{\beta _2}\left\{ \prod _{i_1=1}^{n-1}B_{2i_1+3} \right\} + A_5 B_4 \sum_{n=2}^{\beta _2}\left\{ \prod _{i_1=2}^{n-1}B_{2i_1+3} \right\} \right\} x^{2n+2+\lambda } 
\label{eq:47}
\end{equation}
In (\ref{eq:47}) $y_2^{1,2}(x)$ is sub-power series, for the second square brackets inside curly brackets in sequence $c_n$ including two term of $A_n$'s in (\ref{eq:43}) as $\beta _0$=1 and $\beta _1$=2, for the polynomial case in which makes $B_n$ term terminated.

By using similar process as I did before, the solution for $\beta _1$=3 with $B_8$=0 for the second square brackets inside curly brackets in sequence $c_n$ including $A_2$ in (\ref{eq:43}) is
\begin{eqnarray}
y_2^{1,3}(x) &=& c_0 A_2 B_1 \left\{ A_3 \sum_{n=1}^{\beta _2}\left\{ \prod _{i_1=1}^{n-1}B_{2i_1+3} \right\}+ A_5 B_4 \sum_{n=2}^{\beta _2}\left\{ \prod _{i_1=2}^{n-1}B_{2i_1+3} \right\}\right.\nonumber\\
&&+\left. A_7 B_{4,6} \sum_{n=3}^{\beta _2}\left\{ \prod _{i_1=3}^{n-1}B_{2i_1+3} \right\} \right\} x^{2n+2+\lambda } 
 \label{eq:48}
\end{eqnarray}
By repeating this process for all $\beta _1 =4,5,6,\cdots$, we obtain every $y_2^{1,j}(x)$ terms where $j \geq 4$ for the second square brackets inside curly brackets in sequence $c_n$ including two term of $A_n$'s in (\ref{eq:43}). 
According to (\ref{eq:42}), (\ref{eq:45}), (\ref{eq:47}), (\ref{eq:48}) and every $y_2^{1,j}(x)$ where $j \geq 4$, the general expression of $y_2^1(x)$ for all $\beta _0=1$ replacing the index n by $i_0$ is 
\begin{eqnarray}
y_2^1(x)&=& c_0 \left\{A_0 \sum_{i_0=0}^{\beta _1}\left\{ A_{2i_0+1} \prod _{i_1=0}^{i_0-1}B_{2i_1+2}  \sum_{i_2=i_0}^{\beta _2} \left\{ \prod _{i_3=i_0}^{i_2-1}B_{2i_3+3} \right\}\right\}\right. \label{eq:49}\\
&&+\left. A_2 B_1 \sum_{i_0=1}^{\beta _1}\left\{ A_{2i_0+1} \prod _{i_1=1}^{i_0-1}B_{2i_1+2}  \sum_{i_2=i_0}^{\beta _2} \left\{ \prod _{i_3=i_0}^{i_2-1}B_{2i_3+3} \right\}\right\} \right\} x^{2i_2+2+\lambda } \nonumber
\end{eqnarray}
(c) As $\beta _0$=2, then $B_5$=0 in (\ref{eq:14}).
\begin{equation}
\begin{tabular}{  l  }
  \vspace{2 mm}
  $c_2= A_{0,1} c_0$ \\
  \vspace{2 mm}
  $c_4 = \Big\{ A_0\cdot 1 \Big[ A_1\cdot 1 \cdot B_3+ A_3 B_2\cdot 1 \Big]+ A_2 B_1\Big[ A_3\cdot 1\cdot 1 \Big] \Big\} c_0  $ \\
  \vspace{2 mm}
  $c_6 = \Big\{ A_0 \cdot 1\Big[ A_1\cdot 1 \cdot B_{3,5}+ A_3 B_2 B_5 + A_5 B_{2,4}\cdot 1 \Big] + A_2 B_1 \Big[ A_3\cdot 1\cdot B_5 + A_5 B_4 \cdot 1 \Big] $ \\
  \vspace{2 mm}
  \hspace{0.8 cm} $+ A_4 B_{1,3} \Big[ A_5 \cdot 1\cdot 1 \Big] \Big\} c_0 $\\ 
  \vspace{2 mm}
  $c_8 = \Big\{ A_0\cdot 1 \Big[ A_1\cdot 1 \cdot B_{3,5,7}+ A_3 B_2 B_{5,7}\cdot 1+ A_5 B_{2,4}B_7+A_7 B_{2,4,6}\cdot 1 \Big] $ \\
 \vspace{2 mm}
  \hspace{0.8 cm} $+A_2 B_1 \Big[ A_3\cdot 1\cdot B_{5,7} + A_5 B_4 B_7 + A_7 B_{4,6} \cdot 1 \Big]+ A_4 B_{1,3} \Big[ A_5 \cdot 1\cdot B_7+ A_7 B_6\cdot 1\Big] \Big\} c_0 $\\ 
  \vspace{2 mm}
  $c_{10} = \Big\{ A_0 \cdot 1\Big[ A_1\cdot 1 \cdot B_{3,5,7,9} + A_3 B_2 B_{5,7,9} + A_5 B_{2,4}B_{7,9} +A_7 B_{2,4,6}B_{9} + A_9 B_{2,4,6,8}\cdot 1 \Big] $ \\
 \vspace{2 mm}
  \hspace{0.8 cm} $+A_2 B_1 \Big[ A_3\cdot 1\cdot B_{5,7,9} + A_5 B_4 B_{7,9} + A_7 B_{4,6} B_9  + A_9 B_{4,6,8}\cdot 1 \Big] $\\ 
 \vspace{2 mm}
  \hspace{0.8 cm} $+A_4 B_{1,3} \Big[ A_5\cdot 1\cdot B_{7,9}+ A_7 B_6 B_9+ A_9 B_{6,8}\cdot 1\Big] \Big\} c_0 $\\ 
\hspace{2 mm}
  \large{\vdots}\hspace{7cm}\large{\vdots} \\ 
\end{tabular}
 \label{eq:50}
\end{equation}
By repeating similar process from the above, the general expression of $y_2^2(x)$ for all $\beta _0=2$ in (\ref{eq:50}) is 
\begin{eqnarray}
y_2^2(x)&=& c_0 \left\{A_0 \sum_{i_0=0}^{\beta _1}\left\{ A_{2i_0+1} \prod _{i_1=0}^{i_0-1}B_{2i_1+2}  \sum_{i_2=i_0}^{\beta _2} \left\{ \prod _{i_3=i_0}^{i_2-1}B_{2i_3+3} \right\}\right\}\right. \nonumber\\
&&+ A_2 B_1 \sum_{i_0=1}^{\beta _1}\left\{ A_{2i_0+1} \prod _{i_1=1}^{i_0-1}B_{2i_1+2}  \sum_{i_2=i_0}^{\beta _2} \left\{ \prod _{i_3=i_0}^{i_2-1}B_{2i_3+3} \right\}\right\} \label{eq:51}\\
&&+\left. A_4 B_{1,3} \sum_{i_0=2}^{\beta _1}\left\{ A_{2i_0+1} \prod _{i_1=2}^{i_0-1}B_{2i_1+2}  \sum_{i_2=i_0}^{\beta _2} \left\{ \prod _{i_3=i_0}^{i_2-1}B_{2i_3+3} \right\}\right\} \right\} x^{2i_2+2+\lambda }\nonumber 
\end{eqnarray}
Again by repeating this process for all $\beta _0 =3,4,5,\cdots$, I obtain every $y_2^{j}(x)$ terms where $j \geq 3$.
Then I have general expression $y_2(x)$ for all $\beta _0$ of two term of $A_n$'s according (\ref{eq:42}), (\ref{eq:49}), (\ref{eq:51}) and $y_2^{j}(x)$ terms where $j \geq 3$. 
\begin{eqnarray}
y_2(x)&=& c_0 \sum_{i_0=0}^{\beta _0}\left\{ A_{2i_0} \prod _{i_1=0}^{i_0-1}B_{2i_1+1}  \sum_{i_2=i_0}^{\beta _1} \left\{ A_{2i_2+1} \prod _{i_3=i_0}^{i_2-1}B_{2i_3+2} \sum_{i_4=i_2}^{\beta _2}\left\{ \prod _{i_5=i_2}^{i_4-1}B_{2i_5+3}\right\}  \right\}\right\} x^{2i_4+2+\lambda } \nonumber\\
&&\;\;\mbox{where}\; \beta _0 \leq \beta _1 \leq \beta _2   
\label{eq:52}
\end{eqnarray}
By using similar process for the previous cases of zero, one and two term of $A_n$'s, the function $y_3(x)$ for the case of three term of $A_i$'s is
\begin{eqnarray}
y_3(x) &=& c_0 \sum_{i_0=0}^{\beta _0}\left\{ A_{2i_0} \prod _{i_1=0}^{i_0-1}B_{2i_1+1}  \sum_{i_2=i_0}^{\beta _1} \left\{ A_{2i_2+1} \prod _{i_3=i_0}^{i_2-1}B_{2i_3+2}\right.\right. \label{eq:53}\\
&&\times  \left.\left. \sum_{i_4=i_2}^{\beta _2}\left\{ A_{2i_4+2}\prod _{i_5=i_2}^{i_4-1}B_{2i_5+3} \sum_{i_6=i_4}^{\beta _3} \left\{\prod _{i_7=i_4}^{i_6-1}B_{2i_7+4}  \right\} \right\}  \right\}\right\} x^{2i_6+3+\lambda }\nonumber\\
&&\;\;\mbox{where}\; \beta _0 \leq \beta _1 \leq \beta _2 \leq \beta _3\nonumber  
\end{eqnarray}
By repeating this process for all higher terms of $A_n$'s, I obtain every $y_m(x)$ terms where $m > 3$. Substitute (\ref{eq:25}), (\ref{eq:33}), (\ref{eq:52}), (\ref{eq:53}) and including all $y_m(x)$ terms where $m > 3$ into (\ref{eq:9}). 
\begin{thm}
The general expression of $y(x)$ for the polynomial case which makes $B_n$ term terminated in the three term recurrence relation is
\begin{eqnarray}
y(x) &=& y_0(x)+ y_1(x)+ y_2(x)+y_3(x)+\cdots \nonumber\\
&=& c_0 \left\{ \sum_{i_0=0}^{\beta _0} \left( \prod _{i_1=0}^{i_0-1}B_{2i_1+1} \right) x^{2i_0+\lambda}
+ \sum_{i_0=0}^{\beta _0}\left\{ A_{2i_0} \prod _{i_1=0}^{i_0-1}B_{2i_1+1}  \sum_{i_2=i_0}^{\beta _1} \left( \prod _{i_3=i_0}^{i_2-1}B_{2i_3+2} \right)\right\} x^{2i_2+1+\lambda}  \right. \nonumber\\
&& + \sum_{N=2}^{\infty } \left\{ \sum_{i_0=0}^{\beta _0} \left\{A_{2i_0}\prod _{i_1=0}^{i_0-1} B_{2i_1+1} 
\prod _{k=1}^{N-1} \left( \sum_{i_{2k}= i_{2(k-1)}}^{\beta _k} A_{2i_{2k}+k}\prod _{i_{2k+1}=i_{2(k-1)}}^{i_{2k}-1}B_{2i_{2k+1}+k+1}\right) \right.\right.\nonumber\\
&& \times \left.\left.\left. \sum_{i_{2N} = i_{2(N-1)}}^{\beta _N} \left( \prod _{i_{2N+1}=i_{2(N-1)}}^{i_{2N}-1} B_{2i_{2N+1}+N+1} \right) \right\} \right\} x^{2i_{2N}+N+\lambda}\right\}    
 \label{eq:54}
\end{eqnarray}
\end{thm}
On the above, $\beta _i\leq \beta _j$ only if $i\leq j$ where $i,j,\beta _i, \beta _j \in \mathbb{N}_{0}$ 
\begin{thm}
For infinite series, replacing $\beta _0$,$\beta _1$,$\beta _k$ and $\beta _N$ by $\infty $ in (\ref{eq:54})
\begin{eqnarray}
y(x) &=& y_0(x)+ y_1(x)+ y_2(x)+y_3(x)+\cdots \nonumber\\
&=& c_0 \left\{ \sum_{i_0=0}^{\infty } \left( \prod _{i_1=0}^{i_0-1}B_{2i_1+1} \right) x^{2i_0+\lambda} 
+ \sum_{i_0=0}^{\infty }\left\{ A_{2i_0} \prod _{i_1=0}^{i_0-1}B_{2i_1+1}  \sum_{i_2=i_0}^{\infty } \left( \prod _{i_3=i_0}^{i_2-1}B_{2i_3+2} \right)\right\} x^{2i_2+1+\lambda} \right. \nonumber\\
&& + \sum_{N=2}^{\infty } \left\{ \sum_{i_0=0}^{\infty } \left\{A_{2i_0}\prod _{i_1=0}^{i_0-1} B_{2i_1+1} 
 \prod _{k=1}^{N-1} \left( \sum_{i_{2k}= i_{2(k-1)}}^{\infty } A_{2i_{2k}+k}\prod _{i_{2k+1}=i_{2(k-1)}}^{i_{2k}-1}B_{2i_{2k+1}+k+1}\right) \right.\right.\nonumber\\
&& \times \left.\left.\left. \sum_{i_{2N} = i_{2(N-1)}}^{\infty } \left( \prod _{i_{2N+1}=i_{2(N-1)}}^{i_{2N}-1} B_{2i_{2N+1}+N+1} \right) \right\} \right\} x^{2i_{2N}+N+\lambda}\right\} 
\label{eq:55}
\end{eqnarray}
\end{thm}
(\ref{eq:55}) is exactly equivalent to (\ref{eq:20}). (\ref{eq:55}) is the another general expression of $y(x)$ for the infinite series.
\section{\label{sec:level5} Recurrence relation and Fibonacci numbers of higher order}
\subsection{ Three term recurrence relation and Fibonacci sequence}
The Fibonacci sequence is:
\begin{equation}
1,1,2,3,5,8,13,21,34,55,89,144,\cdots
\label{eq:77}
\end{equation}
The numbers in the sequence follows the recursive relation
\begin{equation}
c_{n+1}= c_n + c_{n-1} \hspace{1cm} : n\geq 1 
\label{eq:78}
\end{equation}
with seed values
\begin{equation}
c_0 = 1\hspace{1cm} c_1 = 1 
\label{eq:79}
\end{equation}
The power series of the generating function of the Fibonacci sequence is 
\begin{equation}
\sum_{n=0}^{\infty } c_n x^n = \frac{x}{1-x-x^2}
\label{eq:80}
\end{equation}
As we know, the three term recurrence relation is
\begin{equation}
c_{n+1}=A_n \;c_n +B_n \;c_{n-1} \hspace{1cm};n\geq 1
\label{eq:81}
\end{equation}
with seed values
\begin{equation}
c_1 = A_0 c_0
\label{eq:82}
\end{equation}
(\ref{eq:7}) is the expansion of (\ref{eq:81}).
If coefficients $A_n= B_n =1$ in (\ref{eq:81}) and $c_0= c_1=1$ in (\ref{eq:82}), then $c_n$ follows Fibonacci sequence. 
Lucas series is another sequence generated by the three term recurrence relation with constant coefficients $A_n= B_n =1$ in (\ref{eq:81}) and $c_0=2, c_1=1$ in (\ref{eq:82}). You can think of these two sequences having constant coefficients as the most basic three term recurrence relation. In contrast, Heun and Mathieu equations coefficients $A_n$ and $B_n$ are defined to be non-constant (second order polynomial in denominator and numerator: see (3), (4a)-(4c) in Ref.\cite{Chou2012c} and (4), (5a)-(5c) in Ref.\cite{ Chou2012e})
The generating function of the Fibonacci sequence corresponds to infinite series of three term recurrence relation. If all coefficients $A_n= B_n =1$ in  (\ref{eq:20}) and (\ref{eq:55}), then it is equivalent to (\ref{eq:80}). 
\subsection{ Two term recurrence relation}
There is an algebraic number sequence in which is
\begin{equation}
1,1,1,1,1,1,\cdots
\label{eq:83}
\end{equation}
I call (\ref{eq:83}) the identity sequence, and it's recurrence relation is
\begin{equation}
c_{n+1}= c_n \hspace{1cm} : n\geq 0 
\label{eq:84}
\end{equation}
with seed values
\begin{equation}
c_0 = 1
\label{eq:85}
\end{equation}
The power series of the generating function of the identity sequence is
\begin{equation}
\sum_{n=0}^{\infty } c_n x^n = \frac{1}{1-x}
\label{eq:86}
\end{equation}
If $B_n$=0 in (\ref{eq:81}), then three term recurrence relation becomes a two term recurrence relation.
\begin{equation}
c_{n+1}=A_n c_n \hspace{1cm};n\geq 0
\label{eq:87}
\end{equation}
Some examples are the Legendre function, Kummer function, hypergeometric function, Bessel function, etc. The number of each sequence $c_n$ in (\ref{eq:87}) is
\begin{equation}
1,1,1,1,1,1,\cdots
\label{eq:88}
\end{equation}
(\ref{eq:88}) is equivalent to (\ref{eq:83}). As I put $A_n$=1 in (\ref{eq:87}), it becomes (\ref{eq:84}). You can think of this sequence having constant coefficients as the most basic two term recurrence relation.
The power series expansion of (\ref{eq:87}) for the infinite series is
\begin{equation}
y(x)= \sum_{n=0}^{\infty } c_n x^{n+\lambda } = c_0 \sum_{n=0}^{\infty } \left( \prod _{i=0}^{n-1}A_{i}\right) x^{n+\lambda }
\label{eq:89}
\end{equation}  
And polynomial case of (\ref{eq:87}) is  
\begin{equation}
y(x)= \sum_{n=0}^{\alpha _0} c_n x^{n+\lambda } = c_0 \sum_{n=0}^{\alpha _0} \left( \prod _{i=0}^{n-1}A_{i}\right) x^{n+\lambda }
\label{eq:90}
\end{equation}
The generating function of the identity sequence corresponds to infinite series of two term recurrence relation. If all coefficients $A_n=1$ in  (\ref{eq:89}), then it definitely equivalent to (\ref{eq:86}). 
\subsection{ Four term recurrence relation and Tribonacci sequence}
Now, let's think about four term recurrence relation in ordinary differential equation. The four term recurrence formula is
\begin{equation}
c_{n+1}=A_n \;c_n +B_n \;c_{n-1}+ C_n c_{n-2} \hspace{1cm};n\geq 2
\label{eq:91}
\end{equation}
with seed values
\begin{equation}
c_1 = A_0 c_0 \hspace{1cm} c_2= \left( A_0 A_1 + B_1 \right) c_0
\label{eq:92}
\end{equation}   
And the number of each of sequence $c_n$ in (\ref{eq:91}) is the following way:
\begin{equation}
1,1,2,4,7,13,24,44,\cdots
\label{eq:93}
\end{equation}
(\ref{eq:93}) is Tribonacci number, and it's recurrence relation is
\begin{equation}
c_{n+1}=c_n +c_{n-1}+ c_{n-2} \hspace{1cm};n\geq 2
\label{eq:94}
\end{equation}
with seed values
\begin{equation}
c_0 = 0\hspace{1cm} c_1= 1 \hspace{1cm} c_2=1
\label{eq:95}
\end{equation}  
If $A_n = B_n = C_n =1$ in (\ref{eq:91}), it's exactly equivalent to Tribonacci recurrence relation. Four term recurrence relation (non-constant coefficients $A_n$, $ B_n$  and $ C_n$) is the more general form than Tribonacci recurrence relation (constant coefficients $A_n$, $ B_n$  and $ C_n$). 
\subsection{ Five term recurrence relation and Tetranacci sequence}
And five term recurrence relation in ordinary differential equation is
\begin{equation}
c_{n+1}=A_n \;c_n +B_n \;c_{n-1}+ C_n c_{n-2}+ D_n c_{n-3} \hspace{1cm};n\geq 3
\label{eq:96}
\end{equation}
with seed values
\begin{equation}
c_1 = A_0 c_0,\hspace{.5cm} c_2= \left( A_0 A_1 + B_1 \right) c_0,\hspace{.5cm} c_3= \left( A_0 A_1 A_2 + A_0 B_2 + A_2 B_1 + C_2\right) c_0
\label{eq:97}
\end{equation}   
And the number of each of sequence $c_n$ in (\ref{eq:96}) is the following way:
\begin{equation}
1,1,2,4,8,15,29,56,108,208,\cdots
\label{eq:98}
\end{equation}
(\ref{eq:98}) is Tetranacci number, and it's recurrence relation is
\begin{equation}
c_{n+1}=c_n +c_{n-1}+ c_{n-2}+ c_{n-3} \hspace{1cm};n\geq 3
\label{eq:99}
\end{equation}
with seed values
\begin{equation}
c_0 = 0\hspace{1cm} c_1= 1 \hspace{1cm} c_2=1\hspace{1cm} c_3=2
\label{eq:100}
\end{equation}  
If $A_n = B_n = C_n =D_n=1$ in (\ref{eq:96}), it becomes exactly equivalent to Tetranacci recurrence relation. From the above multi-term recurrence relation is more general form than n-nacci recurrence relation. 

\section{\label{sec:level6}Conclusion}
In this paper I show how to generalize three-term recurrence relation for polynomials and infinite series analytically. In the next papers I will work out the analytic solution for the three term recurrence relation such as Heun equation and its confluent form, Lame equation, Mathieu equation. I will derive the power series expansion, the integral form and the generating function of Heun, Mathieu, etc functions. (see section 7 for more details) 
\section{\label{sec:level7}Series ``Special functions and three-term recurrence formula (3TRF)''} 

This paper is 2nd out of 10.
\vspace{3mm}

1. ``Approximative solution of the spin free Hamiltonian involving only scalar potential for the $q-\bar{q}$ system'' \cite{Chou2012a} - In order to solve the spin-free Hamiltonian with light quark masses we are led to develop a totally new kind of special function theory in mathematics that generalize all existing theories of confluent hypergeometric types. We call it the Grand Confluent Hypergeometric Function. Our new solution produces previously unknown extra hidden quantum numbers relevant for description of supersymmetry and for generating new mass formulas.
\vspace{3mm}

2. ``Generalization of the three-term recurrence formula and its applications'' \cite{Chou2012b} - Generalize three term recurrence relation in linear differential equation.  Obtain the exact solution of the three term recurrence relation for polynomials and infinite series.
\vspace{3mm}

3. ``The analytic solution for the power series expansion of Heun function'' \cite{Chou2012c} -  Apply three term recurrence formula to the power series expansion in closed forms of Heun function (infinite series and polynomials) including all higher terms of $A_n$'s.
\vspace{3mm}

4. ``Asymptotic behavior of Heun function and its integral formalism'', \cite{Chou2012d} - Apply three term recurrence formula, derive the integral formalism, and analyze the asymptotic behavior of Heun function (including all higher terms of $A_n$'s). 
\vspace{3mm}

5. ``The power series expansion of Mathieu function and its integral formalism'', \cite{Chou2012e} - Apply three term recurrence formula, analyze the power series expansion of Mathieu function and its integral forms.  
\vspace{3mm}

6. ``Lame equation in the algebraic form'' \cite{Chou2012f} - Applying three term recurrence formula, analyze the power series expansion of Lame function in the algebraic form and its integral forms.
\vspace{3mm}

7. ``Power series and integral forms of Lame equation in   Weierstrass's form and its asymptotic behaviors'' \cite{Chou2012g} - Applying three term recurrence formula, derive the power series expansion of Lame function in   Weierstrass's form and its integral forms. 
\vspace{3mm}

8. ``The generating functions of Lame equation in   Weierstrass's form'' \cite{Chou2012h} - Derive the generating functions of Lame function in   Weierstrass's form (including all higher terms of $A_n$'s).  Apply integral forms of Lame functions in   Weierstrass's form.
\vspace{3mm}

9. ``Analytic solution for grand confluent hypergeometric function'' \cite{Chou2012i} - Apply three term recurrence formula, and formulate the exact analytic solution of grand confluent hypergeometric function (including all higher terms of $A_n$'s). Replacing $\mu $ and $\varepsilon \omega $ by 1 and $-q$, transforms the grand confluent hypergeometric function into Biconfluent Heun function.
\vspace{3mm}

10. ``The integral formalism and the generating function of grand confluent hypergeometric function'' \cite{Chou2012j} - Apply three term recurrence formula, and construct an integral formalism and a generating function of grand confluent hypergeometric function (including all higher terms of $A_n$'s). 
\vspace{3mm}

\bibliographystyle{model1a-num-names}
\bibliography{<your-bib-database>}

\end{document}